# Routing Single Photons from a Trapped Ion Using a Photonic Integrated Circuit

Uday Saha [1,2], James D. Siverns[1,2,3], John Hannegan[2,3], Mihika Prabhu[4], Qudsia Quraishi[5,2,6], Dirk Englund[4], Edo Waks[1,2,3,6] *

[1]Department of Electrical and Computer Engineering, University of Maryland College Park, MD, 20742.
[2]Institute for Research in Electronics and Applied Physics (IREAP), University of Maryland, College Park, MD, 20742.
[3]Department of Physics, University of Maryland College Park, MD, 20742.
[4]Department of Electrical Engineering and Computer Science, Massachusetts Institute of Technology, Cambridge, MA, 02139.
[5]United States Army Research Laboratory, Adelphi, MD, 20783.
[6]Joint Quantum Institute (JQI), University of Maryland, College Park, MD, 20742.
*Corresponding Author: edowaks@umd.edu

**Abstract**

**Trapped ions are promising candidates for nodes of a scalable quantum network due to their long-lived qubit coherence times and high-fidelity single and two-qubit gates. Future quantum networks based on trapped ions will require a scalable way to route photons between different nodes. Photonic integrated circuits from fabrication foundries provide a compact solution to this problem. However, these circuits typically operate at telecommunication wavelengths which are incompatible with the strong dipole emissions of trapped ions. In this work, we demonstrate the routing of single photons from a trapped ion using a photonic integrated circuit. We employ quantum frequency conversion to match the emission of the ion to the operating wavelength of a foundry-fabricated silicon nitride photonic integrated circuit, achieving a total transmission of 31±0.9% through the device. Using programmable phase shifters, we switch the single photons between the output channels of the circuit and also demonstrate a 50/50 beam splitting condition. These results constitute an important step towards programmable routing and entanglement distribution in large-scale quantum networks and distributed quantum computers.**

## Introduction

Trapped ions excel as a platform for quantum networking (*1–6*). They exhibit long qubit coherence times (*7*), high-fidelity single- and two-qubit gates (*8, 9*) as well as the ability to generate single photons entangled with their internal qubit states (*3, 4, 10, 11*). Moreover, they can be trapped in compact surface traps (*12, 13*) with integrated waveguides and grating couplers to deliver the light required for cooling and coherent operations (*14–16*) and close-proximity detectors for state detection (*17*). But integrating these discrete memories into a quantum network requires scalable methods to route single photons between different nodes of a network (*1*, *18*).

Integrated photonic devices offer a compact and scalable solution to realize quantum interconnects that can route photons between nodes of a trapped ion quantum network (*19*, *20*). These devices can act as reconfigurable optical cross-connect switches that control the path of photonic qubits within the network in a programmable way (*1*, *18*). However, the use of integrated photonics to route photons from trapped ions has remained a challenge because the emission of trapped ions typically lies in the ultra-violet and visible wavelength regimes (*3*, *10*, *11*, *21*, *22*). This emission is incompatible with the majority of active photonic integrated circuit platforms, which are mainly designed to work at telecommunication wavelengths (*23*, *24*).

In this work, we demonstrate the routing of single photons from a trapped ion using a photonic integrated circuit designed to operate at the telecom C-band. To convert the visible wavelength emission of the ion to the single-mode operating condition of the photonic integrated circuit, we employ a two-stage quantum frequency conversion process (*25*). We actively route photons between different output ports of the circuit with a total transmission of 31±0.9% through the device. We also implement 50/50 beam splitting, an essential operation for entanglement distribution (*1*, *3*, *18*, *26*). Our photonic integrated circuit is fabricated using a commercial foundry, which is compatible with large scaleup (*27*, *28*). These results open new possibilities for quantum networks where single photons distribute entanglement between trapped ion nodes in a programmable way over long distances.

## Results

Figure 1(a) shows the schematic of the photonic integrated circuit along with an optical microscope image of the device. The device consists of a Mach-Zehnder interferometer composed of two 50/50 directional couplers with a programmable internal phase shifter between them. The device was fabricated in the $Si_3N_4$ LioniX TriPleX process with optoelectronic packaging and fiber pigtail. The waveguides in the photonic circuit are double-striped silicon nitride waveguides. A full description of the waveguide geometry and LioniX TriPleX fabrication process can be found in Ref. (*24*). We control the thermo-optic phase shifter using current-driven chromium heaters on the top of the cladding layer above the waveguide.

We first characterize the transmission behavior of the Mach-Zehnder interferometer with a continuous-wave laser operating at a wavelength of 1534 nm. We couple light into the input port and measure the transmission at the output ports 1 and 2 as a function of the current applied to the internal phase shifter (Fig. 1(b)). The current on the internal phase shifter changes the refractive index of the waveguide creating a phase shift between top and bottom waveguides. By sweeping the current supplied to the phase-shifter, we can continuously shift the output light from Port 1 to 2, achieving a near-maximum transmission out of Port 2 at a current of 16.6 mA (the current limit of the heater) (Fig. 1(b)). We achieve a 50/50 splitting ratio with 11.05 mA applied to the internal phase shifter. We define the extinction ratio of each output port as the ratio between their maximum to minimum transmission. Using this definition, the extinction ratios are 10.2 dB and 7.6 dB for Port 1 and Port 2 respectively. These imperfect extinction ratios can be attributed to fabrication imperfections in the directional couplers which result in deviations from an ideal 50/50 splitting.

These imperfections may be improved by replacing the directional couplers with active Mach-Zehnder interferometers (*29*). The sum of the transmission from the two output ports is 31±0.9% for all heater currents. The losses in the device arise from imperfect coupling from the fiber pigtail, waveguide losses, back-reflection, and optical absorption by the metal contact.

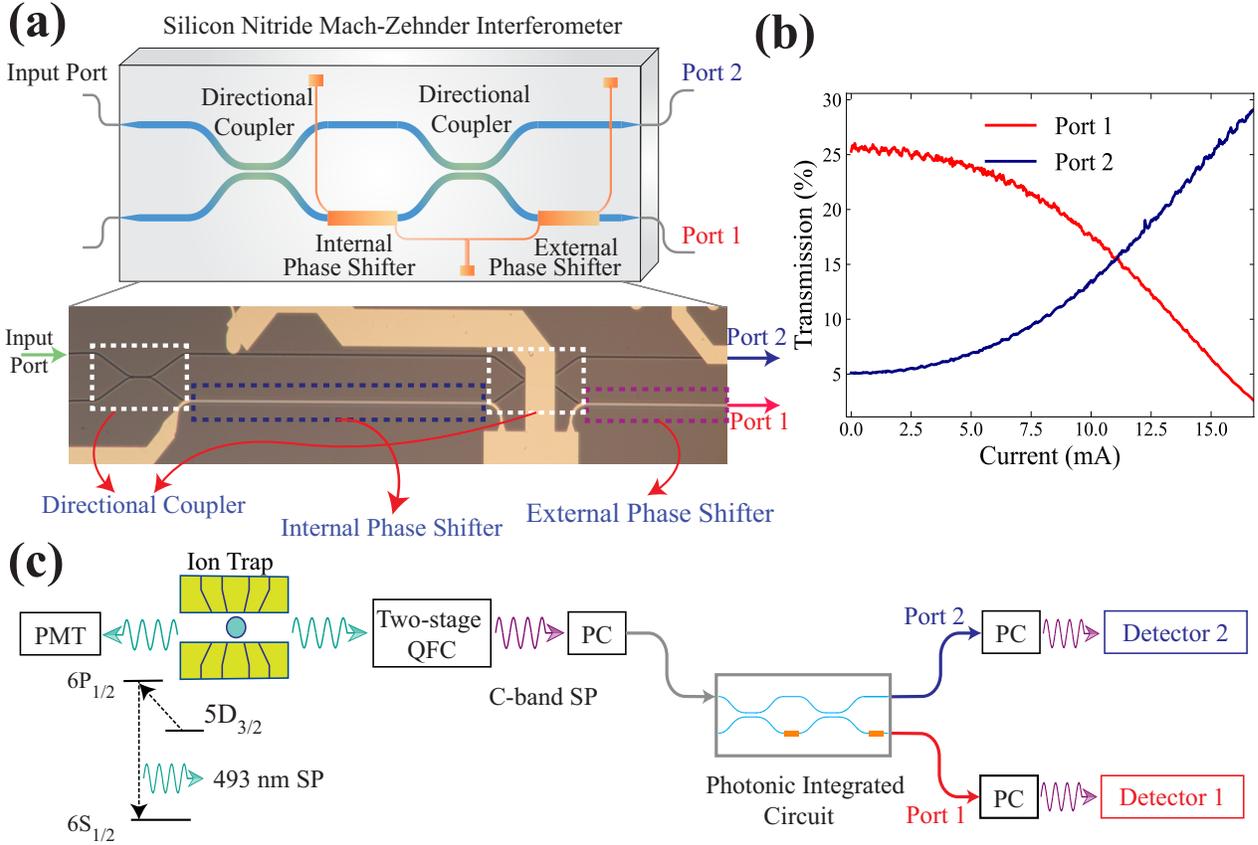

*Fig. 1 Schematic and transmission properties of the Mach-Zehnder interferometer and the experimental layout.*
*(a) A schematic of the silicon nitride Mach-Zehnder interferometer along with an optical microscope image. Highlighted white, blue, and purple boxes are the two directional couplers and the internal and external phase shifters, respectively. (b) The transmission of the Mach-Zehnder interferometer measured at the two output ports as a function of current applied to the internal phase-shifter when laser light is coupled into the input port (c) Illustration of the experimental setup used to route single photons from a trapped barium ion using a silicon nitride photonic integrated circuit. Here PMT, QFC, PC, SP represent photomultiplier tube, quantum frequency conversion, polarization control, and single photons respectively. Also shown is the basic energy level diagram for $^{138}Ba^+$ depicting the excitation and emission of a single photon with a wavelength of 493 nm.*

Figure 1(c) illustrates the experimental setup used to generate single photons from a trapped ion and couple the emission into the photonic integrated circuit. We excite a barium ion ($^{138}Ba^+$) to the $|6P_{1/2}, m_j=+1/2\rangle$ state, which then spontaneously decays to $|6S_{1/2}, m_j= \pm 1/2\rangle$, emitting a 493-nm single photon, as described in prior work (*22*, *25*, *30*). The vacuum chamber, housing the ion trap, has both a front and back window, allowing for photon collection in two directions. Photons collected from the back window are sent to a photomultiplier tube to measure the temporal photon shape of 493-nm photons throughout the experiment. Photons collected from the front window are sent through a polarizer and converted to 1534 nm using a two-stage quantum frequency conversion system as reported in Ref. (*25*). This wavelength is compatible with the single-mode operation of our designed photonic integrated circuit. We then couple converted single photons to the input port of the silicon nitride Mach-Zehnder interferometer. A free-space polarization controller prior to the fiber-coupler aligns the polarization of the photon to the transverse electric guided mode of the waveguide. We use two superconducting-nanowire single-photon detectors to detect photons at the output ports of the Mach-Zehnder interferometer. Because these detectors are polarization-sensitive, we use additional polarization control stages after the output ports to maximize photodetection efficiency.

Figure 2 shows the time-resolved photon counts at the output ports, relative to a trigger pulse synchronized with the excitation laser. All curves are background-subtracted to account for noise photons and detector dark counts. Fig. 2(a) shows the time-resolved photon counts measured at Port 1 when the current of the internal phase shifter is set to 0 mA. The two vertical green dashed lines denote the time window used to measure the photon arrival events. We determine this time window from the temporal shape of the 493 nm photons measured using the photomultiplier tube. We set the window width to be 32 ns, which captures 75% of the total 493-nm photon counts. The red data points in Fig. 2(a) show a photon signal during the window, consistent with the time-resolved 493 nm photon counts shown in Fig. 2(a) (gold squares). Fig. 2(b) shows the same trace for Port 2. In this case, the histogram shows a large reduction in photon probability, indicating that nearly all single photons are routed to Port 1. Fig. 2(c)-(d) show the same measurements at an internal phase shifter current of 16.6 mA, which corresponds to the reverse condition. Here, the majority of photons exit at Port 2, and we observe only a weak signal at Port 1. These measurements demonstrate the routing of single photons from a trapped ion to different output ports of the photonic integrated circuit. Fig. 2(e)-(f) show the time-resolved photon counts at output ports 1 and 2, respectively when we apply a current of 11.05 mA to the internal phase shifter, which corresponds to the 50/50 splitting condition. In this case, we measure the time-resolved photon counts at both ports with similar amplitudes.

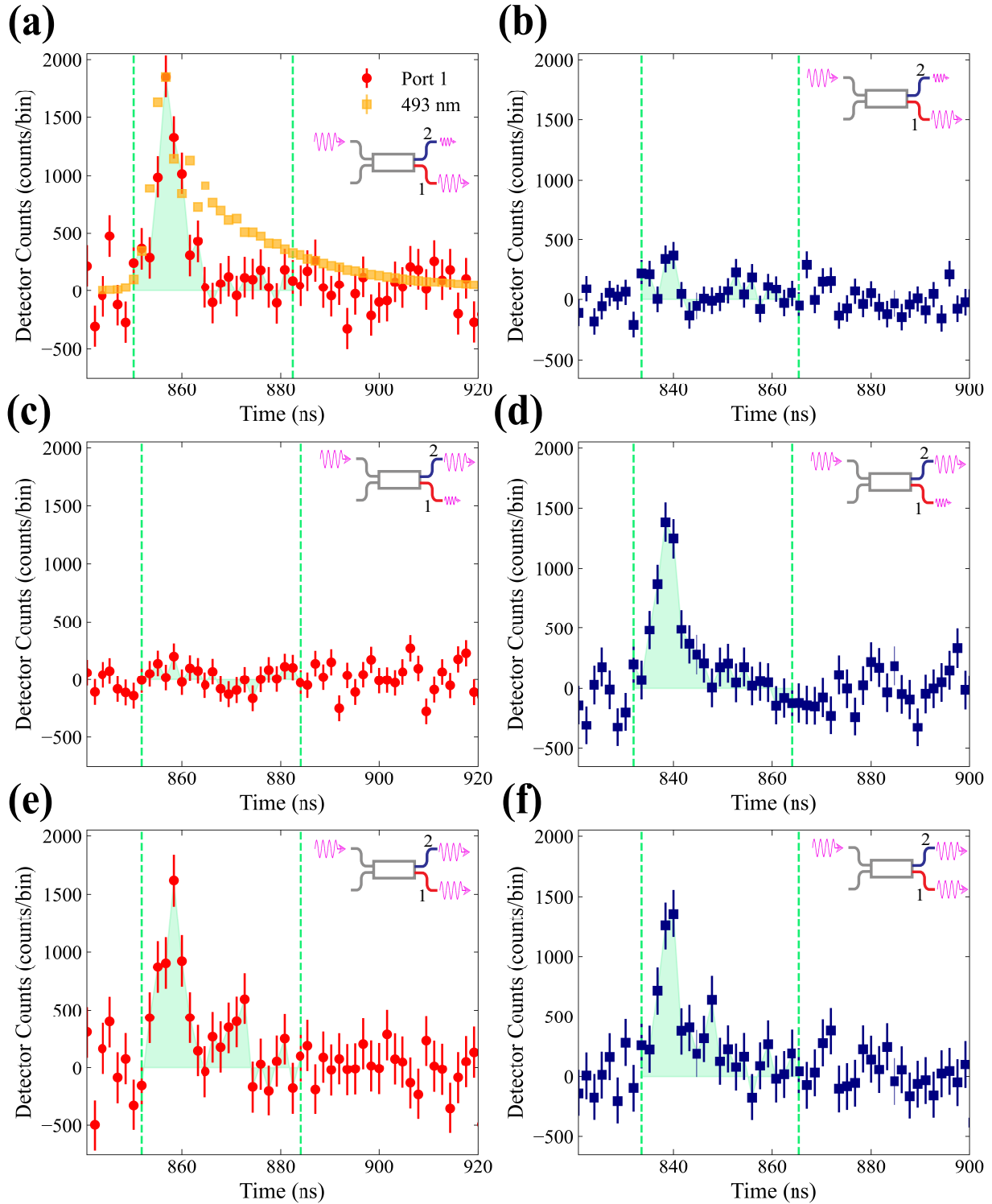

**Fig. 2 The temporal photon shapes at the output ports of the Mach-Zehnder interferometer at different current levels of the internal phase shifter** (a) and (b) are the time-resolved photon counts at Port 1 and Port 2 when the interferometer is set at the maximum transmission of Port 1 (0 mA current on the internal phase shifter). In these plots, the horizontal axis shows the time difference between the arrival of the photon and the trigger pulse. (c) and (d) show the time-resolved photon counts measured at Port 1 and Port 2 when the internal phase shifter current is set to have nearly maximum transmission at Port 2 (16.6 mA current on the internal phase shifter). Finally, (e) and (f)

*are the time-resolved photon counts measured at Port 1 and Port 2 for a 50/50 splitting condition of the Mach-Zehnder interferometer (11.05 mA current on the internal phase shifter). The green vertical lines in all plots indicate the window used to count photon arrival events and correspond to 75% of the total 493 nm time-resolved photon counts (scaled and shown in gold in (a)). The green shaded area represents the photon area measured at each output port. We use a 1.6 ns bin size and perform background subtraction to plot the temporal photon shapes. All errors correspond to shot noise-limited accuracy.*

Figure 3 shows the splitting ratios measured using the photon areas in Fig. 2, superimposed over the measured splitting ratio using a continuous wave laser (red and blue dotted lines). The red circles and blue squares represent the splitting ratios of Port 1 and Port 2 respectively. These values are corrected for mismatch of the photodetection efficiencies between the detectors at Port 1 and Port 2. The ratio of detection efficiencies of the detectors located after Port 1 and Port 2 is 1.13± 0.07 with fiber coupling and polarization control losses taken into account. The error bars correspond to the shot noise-limited accuracy based on the photon counts in each measurement. The splitting ratios calculated from the photon shapes show good agreement with the transmission ratios measured with classical laser light, within experimental uncertainties. These measurements confirm that we are able to route single photons from the ion using the photonic integrated circuit in a programmable manner.

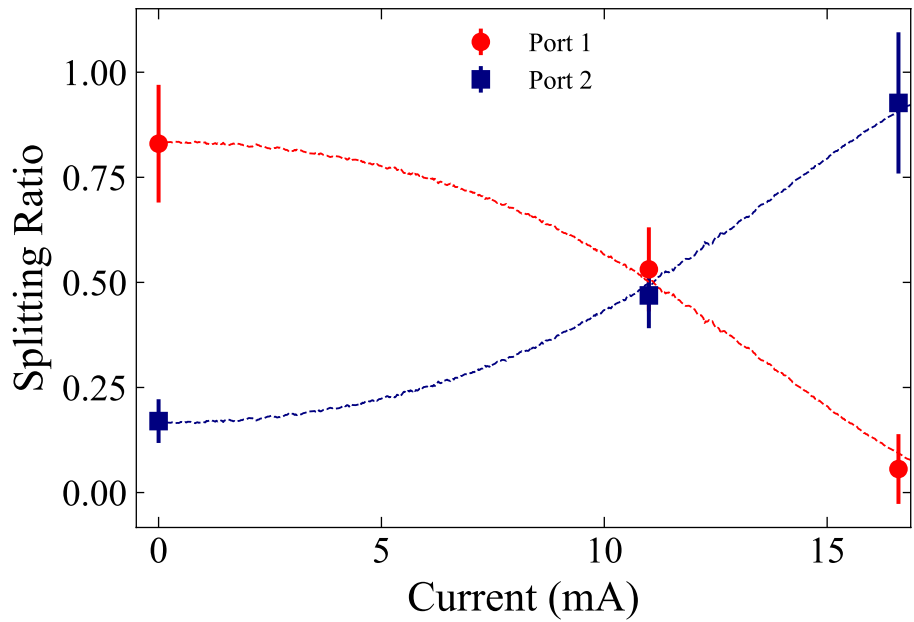

**Fig. 3** *The splitting ratio of the output ports of the Mach-Zehnder interferometer as a function of current applied to the internal phase shifter.* *The data points represent the splitting ratio calculated from the background-corrected time-resolved photon counts in Fig.2. The error bars are calculated from the shot noise of the photon counts. The dotted red and blue lines show the splitting ratio of Port 1 and 2 measured using classical laser light calculated using the data in Fig 1(b).*

## Discussion

In summary, we have demonstrated programmable switching of single photons emitted from a trapped ion between different channels on a photonic integrated chip. Our current proof-of-concept demonstration can be extended in future devices in a straightforward way to include large numbers of Mach-Zehnder interferometers to achieve N×N cross-connects or other arbitrary unitary matrix transformations (*31*, *32*). In addition to routing, photonic integrated circuits can take on other functionalities such as filtering (*33*), wavelength division multiplexing (*34*), and frequency conversion (*35*). They can also implement fundamental quantum operations such as photonic Bell-state analysis (*1*, *3*, *18*, *26*), which is necessary to entangle quantum nodes over long distances. These capabilities could ultimately be used to realize scalable and programmable quantum networks using trapped ion quantum technology.

## Materials and Methods

### Photon Production

To generate visible single photons, we first doppler cool a barium ion and then optically pump it into the $|5D_{3/2}, m_j=3/2\rangle$ state using $\pi$ and $\sigma-$ polarized 650 nm light and $\pi$ polarized 493 nm light (*6*, *22*, *25*, *30*). We then excite the ion using $\sigma+$ polarized 650 nm light to the $|6P_{1/2}, m_j=+1/2\rangle$ state which spontaneously decays to the $|6S_{1/2}, m_j=\pm1/2\rangle$ ground state, emitting a 493-nm single photon (Fig. 4). The repetition rate of photon production attempts is 780.64 kHz. In each cycle of photon production, we wait for 320 ns, directly after optical pumping, with all the lasers being off, to ensure that no stray light is detected during the photon extraction process.

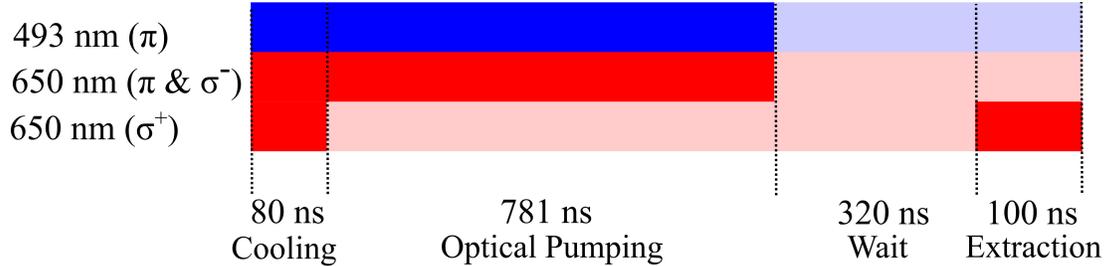

**Fig. 4 Experiment timing sequence**. *The experimental timing sequence that we use to cool, optically pump and extract single 493 nm photons from the trapped barium ion.*

### Quantum Frequency Conversion

We use a two-stage quantum frequency conversion scheme to generate telecom C-band photons from the ion (*25*). In the first stage of quantum frequency conversion, we generate 780 nm single photons from 493 nm single photons utilizing a 1343 nm pump laser with a conversion efficiency of 20% (*36*, *37*). Then, in a second stage of conversion, we convert 780 nm single photons to 1534 nm single photons using a 1589 nm pump laser (*25*). We achieve 25% conversion-efficiency in the second stage of quantum frequency conversion, including an ~80% efficient 20 GHz bandwidth tunable filter and ~60% transmissive etalon of 46.1 MHz bandwidth. In both frequency conversion processes, we use a difference frequency generation scheme with periodically poled lithium-niobate waveguides (*25*, *36*).

### Frequency conversion pump laser locking

The frequency of the converted photons is determined by the frequencies of the two pump lasers used. To stabilize the frequency of the converted photon, we lock the two pump laser frequencies and ensure that the converted photons do not drift outside the bandwidth of etalon. To lock the pump lasers, we first lock a 1762 nm laser to an ultra-low expansion cavity via a Pound–Drever–Hall technique. This locked laser is then used as a reference laser with which we can lock the conversion pump lasers. We lock the pump lasers relative to the 1762 nm laser using a Fabry-Perot scanning cavity with a scan frequency of 170 Hz resulting in a drift lock of a few MHz (*38*).

## Acknowledgments

We acknowledge the support from LioniX for fabricating the photonic integrated circuit.

All part numbers and company references are given for technical purposes, and their mention does not represent an endorsement on the part of the U.S. government. Other equivalent or better options may be available.

## Conflicts of interest

All authors declare no conflict of financial interest.